\documentclass[reqno]{article}
\RequirePackage[reqno]{amsmath}
\begin{document}
\title{\textbf{Geometric Representation of the  generator of duality in  massless and massive $p-$form field theories}}
\author{ Ernesto Contreras$^{1}$, Lorenzo Leal$^{1,2}$\ and Yisely Martinez$^{1}$\\
\small \textsl{$1.$ Grupo de Campos y Part\'iculas, Facultad de Ciencias,
Universidad Central de Venezuela, AP $47270$,}\\
\small \textsl{Caracas $1041-A$, Venezuela.}\\
\small\textsl{$2.$ Departamento de F\'isica,Universidad Sim\'on Bolivar, AP $89000$,}\\
\small\textsl{Caracas $1080-A$}, Venezuela.}
\date{}
\maketitle

\begin{abstract}
We study the invariance under duality transformations in massless
and massive $p-$form  field theories and obtain the  Noether
generators of the infinitesimal transformations that correspond to
this symmetry. These generators are realized in geometrical
representations that generalize the Loop Representation of the
Maxwell field, allowing for a geometrical interpretation which is
studied.
\end{abstract}

\section{INTRODUCTION}

The duality electricity-magnetism that the Maxwell theory exhibits is still a source of inspiration in the search
and study of similar transformations in non-Abelian gauge theories, statistical models,  string and brane theories and supergravity, among other models. Particular mention deserves the celebrated gauge/string duality \cite{maldacena}, that sheds light in the understanding of the non perturbative behavior of gauge theories, which is a mayor challenge of theoretical physics.
Apart of its usefulness, duality is a very appealing symmetry: it relates seemingly unrelated models, or switches
 between different aspects of the same model in a sometimes unexpected fashion.

As it was shown in reference \cite{deser}, the electric-magnetic duality of the Maxwell theory can be seen as a
symmetry of both the equations of motion and the action when the duality transformation is implemented in terms of
 the Lagrangian variables $A_{\mu}$, in the second order or Lagrangian formulation, or in terms of the variables
 $A_{\mu}$ and their canonical conjugates in the Hamiltonian or first order formulation. From this point of view,
 the electric-magnetic duality gives rise to a Noether current, whose associated charge generates the duality rotations
  of the variables of the theory. More precisely, the $SO(2)$ duality rotations

\[
\left(\begin{array}{c}
 E'^{i}\\
B'^{i} \end{array}\right)= \left(\begin{array}{cc}
cos\theta&sen\theta\\
-sen\theta&cos\theta\end{array}\right)
 \left(\begin{array}{c}
 E^{i}\\
B^{i} \end{array}\right)
\]

that interchange electric and magnetic fields, or, equivalently,
$F^{\mu\nu}$ and its dual $*F^{\mu\nu}$, can be derived from the infinitesimal transformations

\begin{eqnarray*}
\delta{\vec{E}}&=&\beta\nabla\times\vec{A}\\
\delta{\vec{A}}&=&\beta\nabla^{-2}\nabla\times\vec{E}
\end{eqnarray*}
 which involve the canonical variables of the theory. It can be seen that the action $S=-\frac{1}{4}\int d^{4} F_{\mu\nu}F^{\mu\nu}$, when written in its first order form, and the equations of motion, are invariant
under these infinitesimal  transformations (indeed, the action changes by a total time derivative) \cite{deser}. The constant of motion obtained from Noether theorem that generates this duality transformation results to be
$G=-\frac{1}{2}\int
d^{3}x(-A^{i}\varepsilon^{ijk}\partial_{j}A^{k}+E^{i}\nabla^{-2}\varepsilon^{ijk}\partial_{j}E^{k})$.

Models with $p$-form  fields, which constitute a generalization of Maxwell theory or Proca theory (depending on wether they are massive or massless), also exhibit a duality symmetry when the dimension of space time and the rank $p$ are appropriately related. These models play an important role in  superstring and supergravity theories \cite{green}. Furthermore, $p$-forms couple in a natural way to $(p-1)$-branes, in a straightforward generalization of the coupling of electric or magnetic point charges with the Maxwell field.

In $D$ dimensions, massless theories of ranks $p$, $q$, with $p + q = D-2$ are mutually dual. For
example, in four dimensions, Maxwell theory is self-dual, while the second
rank gauge theory is dual to the massles scalar field. Regarding the massive cases, it is found that duality is obtained when the ranks of the forms
obey $p + q = D-1$. For instance, the Proca model is self dual in $D=3$.

The realization of electric-magnetic duality as a symmetry of the
action, developed in \cite{deser} for the Maxwell field, can be
also extended to the massless or massive theories of $p$-forms of
arbitrary rank mentioned above \cite{deserteitel1, deserteitel2,
henne, wotza, nor, mene}. It is found that the duality group is
$Z_{2}$ or $SO(2)$ (in the last case it is then a Noether
symmetry) depending on the relationship between the dimension and
the rank of the forms involved and on the massive or massless
character of the model (see tables (2) and (3)  of reference
\cite{mene}).

In this article, after revisiting the canonical formulation of
antisymmetric field theories, we calculate the generator of
$SO(2)$ duality transformations for the massless and massive
situations, both in the cases of self-duality (i.e., when the theory
becomes mapped onto itself under duality rotations) and of duality
among two different theories. Then, the theories are quantized in
certain geometrical representations that generalize the
Loop Representation of the Maxwell field \cite{gambtri,
gambpull,dibargambtri,jaime} in an appropriate manner. The generators of
duality are also realized in these  geometrical
representation. It is found that the action of the generators onto
the wave functionals captures topological invariants of the
geometric arguments (the generalized Loops) of these functionals.
These invariants are generalized linking numbers and numbers of
intersection among $p$-surfaces. This result is a consequence of
the fact that the generator of duality is metric-independent, as
we shall discuss.

The paper is organized as follows. In sections $2$ and $3$ we
study the canonical quantization and find the generator of duality
of massless and massive models of $p$-forms. In section $4$ we
discuss the path and loop representations  \cite{gambtri,
gambpull,dibargambtri,jaime}, present a generalization of this
representation to $p$-surfaces \cite{piogae,pioernesyo}, and discuss the
realization of the generator within this geometrical setting.
\\

\section{MASSLESS MODELS}
\subsection{Self Dual Massless Models}
The action for the theory of a free massless  $p-$form field in  $D$-dimensional flat space-time
can be taken as
\begin{eqnarray}
S=\frac{(-1)^{p}}{2(p+1)!}\int d^{D}x
F_{\mu_{1}...\mu_{p+1}}F^{\mu_{1}...\mu_{p+1}}\label{anmg},
\end{eqnarray}
where $F_{\mu_{1}...\mu_{p+1}}$ is the field strength
defined by
\begin{eqnarray}
F_{\mu\mu_{1}...\mu_{p}}=\frac{1}{p!}\partial_{[\mu}A_{\mu_{1}...\mu_{p}]}\label{fmn},
\end{eqnarray}
and the $p-$form $A_{\mu_{1}...\mu_{p}}$ is the antisymmetric potential. The equations of motion are given by
\begin{eqnarray}
\partial_{\mu}F^{\mu\mu_{1}...\mu_{p}}=0 \label{emml}.
\end{eqnarray}
\\

When $D-2=2p$ the theory is self-dual, since equation
(\ref{fmn}) can be written down as
\begin{eqnarray}
\partial_{\mu}*F^{\mu\mu_{1}...\mu_{p}}=0,\label{emmlh}
\end{eqnarray}
and the dual field strength $*F^{\mu\mu_{1}...\mu_{p}}=
\frac{1}{(p+1)!}\varepsilon^{\mu\mu_{1}...\mu_{p}\nu_{1}...\nu_{p+1}}F_{\nu_{1}...\nu_{p+1}}$
is also a $(p+1)-$form. To display the duality symmetry in terms of
the generalized electric and magnetic fields
$E^{i_{1}...i_{p}}=F^{i_{1}...i_{p}0}$ and
$B^{i_{1}...i_{p}}=\frac{(-1)^{p}}{(p+1)!}\varepsilon^{i_{1}...i_{p}j_{1}...j_{p+1}}F_{j_{1}...j_{p+1}}$,
we expand equations (\ref{emml}) and (\ref{emmlh}) in the form
\begin{eqnarray}\label{emg}
\partial_{i}E^{ii_{1}...i_{p-1}}=0,\nonumber\\
-\partial_{0}E^{i_{1}...i_{p}}+\frac{1}{p!}\varepsilon^{i_{1}...i_{p}jj_{1}...j_{p}}\partial_{j} B^{j_{1}...j_{p}}=0,\nonumber\\
\partial_{j}B^{jj_{1}...j_{p-1}}=0,\nonumber\\
(-1)^{p}\partial_{0}B^{i_{1}...i_{p}}-\frac{1}{p!}\varepsilon^{i_{1}...i_{p}j_{1}...j_{p}}\partial_{j}E^{j_{1}...j_{p}}=0.
\end{eqnarray}

 As it was studied in references \cite{deser,deserteitel1,deserteitel2,mene},
when $p$ is even the equations of motion are invariant under the
$Z_{2}$ transformations
\begin{eqnarray}
\left(\begin{array}{c}
 E'^{i_{1}...i_{p}}\\
B'^{i_{1}...i_{p}} \end{array}\right)= \left(\begin{array}{cc}
0&1\\
1&0\end{array}\right)
 \left(\begin{array}{c}
 E^{i_{1}...i_{p}}\\
B^{i_{1}...i_{p}} \end{array}\right),
\end{eqnarray}
whereas if $p$ is odd, the equations of motion are left invariant under
$SO(2)$ duality rotations
\begin{eqnarray}\label{ebbe}
\left(\begin{array}{c}
 E'^{i_{1}...i_{p}}\\
B'^{i_{1}...i_{p}} \end{array}\right)= \left(\begin{array}{cc}
cos\theta&sen\theta\\
-sen\theta&cos\theta\end{array}\right)
 \left(\begin{array}{c}
 E^{i_{1}...i_{p}}\\
B^{i_{1}...i_{p}} \end{array}\right).
\end{eqnarray}
\\

In the spirit of reference \cite{deser}, we are interested in the
study of the invariance under duality transformations within the
Hamiltonian framework. Hence, we must carry out the canonical
analysis of the theory. Following  Dirac's procedure for dealing
with singular Lagrangeans \cite{dirac,hene} we obtain the
constraints

\begin{eqnarray}
\phi_{1}&=&E^{i_{1}...i_{p-1}0}=0,\nonumber\\
\phi_{2}&=&\partial_{i_{1}}E^{i_{1}...i_{p}}=0,\nonumber\\
\phi_{3}&=&A_{i_{1}...i_{p-1}0}=0,\nonumber\\
\phi_{4}&=&\partial_{i_{1}}A_{i_{1}...i_{p}}=0.
\end{eqnarray}
The first and second constraints are first class ones, arising from the very definition of the momenta and from the time preservation of the vanishing momenta. The third and fourth are gauge-fixing constraints. Hence, the whole
list  constitutes a set of second-class constraints.

We are interested in the cases where the duality symmetry is continuous, which correspond to $p$ odd. In this case, the first order action results to be
\begin{eqnarray}
S&=&\int
d^{D}x(\frac{1}{p!}E^{i_{1}...i_{p}}\dot{A}_{i_{1}...i_{p}}-\frac{1}{2p!}E^{i_{1}...i_{p}}E^{i_{1}...i_{p}}\nonumber\\
&-&\frac{1}{2(p+1)!}F_{i_{1}...i_{p+1}}F^{i_{1}...i_{p+1}}) .
\end{eqnarray}
It can be verified that varying  this action we obtain the canonical equations of motion, in terms of the gauge-fixed canonical variables.

The  Dirac brackets among the canonical variables are given by
\begin{eqnarray}
\{A_{i_{1}...i_{p}}(\vec{x}),E^{j_{1}...j_{p}}(\vec{y})\}^{*}&=&\delta^{j_{1}...j_{p}}_{i_{1}...i_{p}}\delta^{D-1}(\vec{x}-\vec{y})-(p!)^{2}
\partial^{\vec{x}}_{i_{1}}\partial^{\vec{y}}_{j_{1}}(\delta^{aj_{2}...j_{p}}_{bi_{2}...i_{p}}\partial_{a}\partial_{b})^{-1}
\delta^{D-1}(\vec{x}-\vec{y}),\nonumber\\
\{A_{i_{1}...i_{p}}(\vec{x}),A_{j_{1}...j_{p}}(\vec{y})\}^{*}&=&\{E^{i_{1}...j_{p}}(\vec{x}),E^{j_{1}...j_{p}}(\vec{y})\}^{*}=0,
\end{eqnarray}
\\
where $\delta^{j_{1}...j_{p}}_{i_{1}...i_{p}}$ is the antisymmetric
Kronecker delta, whose value is $+1$ if $j_{1}...j_{p}$ is an even permutation of
$i_{1}...i_{p}$, $-1$ if $j_{1}...j_{p}$ is an odd permutation of
$i_{1}...i_{p}$ and $0$ in the remaining cases. The left hand side of the expression for the bracket among the potential $A$ and its momentum, generalizes the transverse Dirac delta that appears in the canonical formulation of Maxwell theory in the Coulomb gauge, as it should be.

When $p$ is odd, the infinitesimal $SO(2)$ transformations corresponding to equation
(\ref{ebbe}) can be written down in terms of the
canonical variables $A_{i_{1}...i_{p}}$ and $E^{i_{1}...i_{p}}$ as
\begin{eqnarray}\label{potcam}
\delta E^{i_{1}...i_{p}}&=&-\frac{\theta}{p!}\varepsilon^{i_{1}...i_{p}jj_{1}...j_{p}}\partial_{j}A_{j_{1}...j_{p}},\nonumber\\
\delta
A_{i_{1}...i_{p}}&=&-\frac{\theta}{p!}\nabla^{-2}\varepsilon^{i_{1}...i_{p}jj_{1}...j_{p}}\partial_{j}E^{j_{1}...j_{p}}.
\end{eqnarray}
\\

It can be seen that the transformations (\ref{potcam}) change the
first order action by a total time derivative, just as in the usual
Maxwell theory \cite{deser}. Then, Noether theorem allows us to
obtain the generator of infinitesimal duality-rotations , which results
to be
\begin{eqnarray}\label{genoma}
G&=&\frac{1}{2p!^{2}}\int
d^{D-1}x(A_{i_{1}...i_{p}}\varepsilon^{i_{1}...i_{p}jj_{1}...j_{p}}\partial_{j}A_{j_{1}...j_{p}}\nonumber\\
&-&E^{i_{1}...i_{p}}\nabla^{-2}\varepsilon^{i_{1}...i_{p}jj_{1}...j_{p}}\partial_{j}E^{j_{1}...j_{p}}).
\end{eqnarray}
\\

This expression for the generator includes the four dimensional Maxwell case, which was studied by Deser and
Teitelboim \cite{deser}. In that case, $p=1$ and we have
\begin{eqnarray}\label{gnmad}
G=\frac{1}{2}\int d^{3}x(A^{i}\varepsilon^{ijk}\partial_{j}A^{k}
-E^{i}\nabla^{-2}\varepsilon^{ijk}\partial_{j}E^{k}).
\end{eqnarray}

We shall come back to the four dimensional Maxwell theory when discussing the realization of the theories in the Loop Representation and its generalizations. The Maxwell case will give us some insight about the geometrical interpretation of the generator that will be useful for the other cases under study .

To quantize the theory we promote the canonical variables to
operators obeying equal time canonical commutators
\begin{eqnarray}\label{algae}
\left[\hat{A}_{i_{1}...i_{p}}(\vec{x}),\hat{E}^{j_{1}...j_{p}}(\vec{y})\right]&=&i\delta^{j_{1}...j_{p}}_{i_{1}...i_{p}}\delta^{D-1}(\vec{x}-\vec{y})
-i(p!)^{2}\partial^{x}_{i_{1}}\partial^{y}_{j_{1}}(\delta^{aj_{2}...j_{p}}_{bi_{2}...i_{p}}\partial_{a}\partial_{b})^{-1}\delta^{D-1}(\vec{x}-\vec{y}),\nonumber\\
\left[\hat{A}_{i_{1}...i_{p}}(\vec{x}),\hat{A}_{j_{1}...j_{p}}(\vec{y})\right]&=&\left[\hat{E}^{i_{1}...i_{p}}(\vec{x}),\hat{E}^{j_{1}...j_{p}}(\vec{y})\right]=0,
\end{eqnarray}
and take the evolution of physical states to be given by the
Schroedinger equation
\begin{eqnarray}
i\frac{\partial|\Psi\rangle}{\partial t}=\hat{H}|\Psi\rangle,
\end{eqnarray}
where $\hat{H}=\frac{1}{2p!}\int
d^{D-1}x(\hat{E}^{i_{1}...i_{p}}\hat{E}^{i_{1}...i_{p}}+\hat{B}^{i_{1}...i_{p}}\hat{B}^{i_{1}...i_{p}})$
is the Hamiltonian.
\\

\subsection{Dual Massless Models}
In this section we also consider massless theories, but this time we take the action  as
\begin{eqnarray}\label{tnmdd}
S=\frac{(-1)^{p}}{2(p+1)!}\int d^{D}x
F_{\mu_{1}...\mu_{p+1}}F^{\mu_{1}...\mu_{p+1}}+\frac{(-1)^{q}}{2(q+1)!}\int
d^{D}x G_{\mu_{1}...\mu_{q+1}}G^{\mu_{1}...\mu_{q+1}},
\end{eqnarray}
where
\begin{eqnarray}\label{fmnd}
F_{\mu\mu_{1}...\mu_{p}}=\frac{1}{p!}\partial_{[\mu}A_{\mu_{1}...\mu_{p}]},\nonumber\\
G_{\mu\mu_{1}...\mu_{q}}=\frac{1}{q!}\partial_{[\mu}C_{\mu_{1}...\mu_{q}]},
\end{eqnarray}
with $A_{\mu_{1}...\mu_{p}}$ and $C_{\mu_{1}...\mu_{q}}$
antisymmetric potentials. The equations of motion are given by
\begin{eqnarray}\label{emfgnm}
\partial_{\mu}F^{\mu\mu_{1}...\mu_{p}}=0,\nonumber\\
\partial_{\mu}G^{\mu\mu_{1}...\mu_{q}}=0.
\end{eqnarray}

The action (\ref{tnmdd}) corresponds to  the theory of two
massless and uncoupled antisymmetric gauge potentials (a $p$ and a
$q$-form). In references \cite{wotza, nor,mene} these potentials
are seen as the real and imaginary parts of a complex gauge field.
The interest in these models rests in the fact that, under certain
circumstances, the two intervening theories map into each other
when we perform duality transformations. More precisely, for
$D-2=p+q$, equations (\ref{fmnd}) can be written down as
\begin{eqnarray}\label{fgdd}
\partial_{\mu}*F^{\mu\mu_{1}...\mu_{q}}=0,\nonumber\\
\partial_{\mu}*G^{\mu\mu_{1}...\mu_{p}}=0,
\end{eqnarray}
showing, together with equations (\ref{emfgnm}), the duality between the $F$ and $G$ fields.

As before, to study the duality transformations in terms of the
generalized electric fields $E^{i_{1}...i_{p}}=F^{i_{1}...i_{p}0}$
and $D^{i_{1}...i_{q}}=G^{i_{1}...i_{q}0}$ and generalized magnetic
fields
$B^{i_{1}...i_{q}}=\frac{(-1)^{p}}{(p+1)!}\varepsilon^{i_{1}...i_{q}j_{1}...j_{p+1}}F_{j_{1}...j_{p+1}}$
and
$H^{i_{1}...i_{p}}=\frac{(-1)^{q}}{(q+1)!}\varepsilon^{i_{1}...i_{p}j_{1}...j_{q+1}}G_{j_{1}...j_{q+1}}$,
we expand equations (\ref{emfgnm}) and (\ref{fgdd}) to obtain
\begin{eqnarray}\label{emg}
\partial_{i}E^{ii_{1}...i_{p-1}}=0,\nonumber\\
-\partial_{0}E^{i_{1}...i_{p}}+\frac{(-1)^{(p+1)(q+2)}}{q!}\varepsilon^{i_{1}...i_{p}jj_{1}...j_{q}}\partial_{j} B^{j_{1}...j_{q}}=0,\nonumber\\
\partial_{j}B^{jj_{1}...j_{p-1}}=0,\nonumber\\
(-1)^{p+1}\partial_{0}B^{i_{1}...i_{q}}+\frac{1}{p!}\varepsilon^{i_{1}...i_{q}jj_{1}...j_{p}}\partial_{j}E^{j_{1}...j_{p}}=0,\nonumber\\
\partial_{i}D^{ii_{1}...i_{q-1}}=0,\nonumber\\
-\partial_{0}D^{i_{1}...i_{q}}+\frac{(-1)^{(q+1)(p+2)}}{p!}\varepsilon^{i_{1}...i_{q}jj_{1}...j_{p}}\partial_{j} H^{j_{1}...j_{p}}=0,\nonumber\\
\partial_{j}H^{jj_{1}...j_{q-1}}=0,\nonumber\\
(-1)^{q+1}\partial_{0}H^{i_{1}...i_{p}}+\frac{1}{q!}\varepsilon^{i_{1}...i_{p}jj_{1}...j_{q}}\partial_{j}D^{j_{1}...j_{q}}=0.
\end{eqnarray}
\\

When $p$ and $q$ are even, these equations of motion are invariant
under the $Z_{2}$ transformations
\begin{eqnarray}
\left(\begin{array}{c}
E'^{i_{1}...i_{p}}\\
H'^{i_{1}...i_{p}}\\
\end{array}\right)= \left(\begin{array}{cc}
0&1\\
1&0\\
\end{array}\right)
\left(\begin{array}{c}
E^{i_{1}...i_{p}}\\
H^{i_{1}...i_{p}}
\end{array}\right),
\end{eqnarray}
and
\begin{eqnarray}
\left(\begin{array}{c}
B'^{i_{1}...i_{p}}\\
D'^{i_{1}...i_{p}}\\
\end{array}\right)= \left(\begin{array}{cc}
0&1\\
1&0\\
\end{array}\right)
\left(\begin{array}{c}
B^{i_{1}...i_{p}}\\
D^{i_{1}...i_{p}}
\end{array}\right).
\end{eqnarray}

On the other hand, if $p$ is odd and $q$ is even or odd  the
equations of motion remain invariant under $SO(2)$ transformations
\begin{eqnarray}
\left(\begin{array}{c}
E'^{i_{1}...i_{p}}\\
H'^{i_{1}...i_{p}}
\end{array}\right)= \left(\begin{array}{cc}
cos\theta&sen\theta\\
-sen\theta&cos\theta\\
\end{array}\right)
\left(\begin{array}{c}
E^{i_{1}...i_{p}}\\
H^{i_{1}...i_{p}}
\end{array}\right),
\end{eqnarray}
and
\begin{eqnarray} \left(\begin{array}{c}
B'^{i_{1}...i_{q}}\\
D'^{i_{1}...i_{q}}
\end{array}\right)= \left(\begin{array}{cc}
cos\theta&(-1)^{q}sen\theta\\
(-1)^{q+1}sen\theta&cos\theta\\
\end{array}\right)
\left(\begin{array}{c}
B^{i_{1}...i_{q}}\\
D^{i_{1}...i_{q}}
\end{array}\right).
\end{eqnarray}
\\

These results agree with those obtained in reference \cite{mene}.

The canonical analysis leads to the following set of first-class constraints and gauge-fixing constraints

\begin{eqnarray}
\phi_{1}=E^{i_{1}...i_{p-1}0}=0\hspace{1cm}                &\phi_{5}=A_{i_{1}...i_{p-1}0}=0\nonumber\\
\phi_{2}=D^{i_{1}...i_{q-1}0}=0\hspace{1cm}                 &\phi_{6}=C_{i_{1}...i_{q-1}0}=0\nonumber\\
\phi_{3}=\partial_{i_{1}}E^{i_{1}...i_{p}}=0\hspace{1cm}    &\phi_{7}=\partial_{i_{1}}A_{i_{1}...i_{p}}=0\nonumber\\
\phi_{4}=\partial_{i_{1}}D^{i_{1}...i_{q}}=0\hspace{1cm}
&\phi_{8}=\partial_{i_{1}}C_{i_{1}...i_{q}}=0.
\end{eqnarray}

Together, they constitute a  set of second-class constraints. When
$p$ is odd and $q$ is even or odd the first order action is
\begin{eqnarray}
S=\int
d^{D}x(\frac{1}{p!}E^{i_{1}...i_{p}}\dot{A}_{i_{1}...i_{p}}+\frac{1}{q!}D^{i_{1}...i_{q}}\dot{C}_{i_{1}...i_{q}}
-\frac{1}{2p!}E^{i_{1}...i_{p}}E^{i_{1}...i_{p}}-\frac{(-1)^{q+1}}{2p!}D^{i_{1}...i_{q}}D^{i_{1}...i_{q}}\nonumber\\
-\frac{1}{2(p+1)!}F_{i_{1}...i_{p+1}}F^{i_{1}...i_{p+1}}+\frac{(-1)^{q}}{2(q+1)!}G_{i_{1}...i_{q+1}}G^{i_{1}...i_{q+1}}).
\end{eqnarray}
The Dirac brackets of the canonical variables are given by
\begin{eqnarray}
\{A_{i_{1}...i_{p}}(\vec{x}),E^{j_{1...j_{p}}}(\vec{y})\}^{*}&=&\delta^{j_{1}...j_{p}}_{i_{1}...i_{p}}\delta^{D-1}(\vec{x}-\vec{y})-(p!)^{2}
\partial^{x}_{i_{1}}\partial^{y}_{j_{1}}(\delta^{aj_{2}...j_{p}}_{bi_{2}...i_{p}}\partial_{a}\partial_{b})^{-1}
\delta^{D-1}(\vec{x}-\vec{y}),\nonumber\\
\{A_{i_{1}...i_{p}}(\vec{x}),A_{j_{1}...j_{p}}(\vec{y})\}^{*}&=&\{E^{i_{1}...j_{p}}(\vec{x}),E^{j_{1}...j_{p}}(\vec{y})\}^{*}=0,\nonumber\\
\{C_{i_{1}...i_{q}}(\vec{x}),D^{j_{1...j_{q}}}(\vec{y})\}^{*}&=&\delta^{j_{1}...j_{q}}_{i_{1}...i_{q}}\delta^{D-1}(\vec{x}-\vec{y})-(q!)^{2}
\partial^{x}_{i_{1}}\partial^{y}_{j_{1}}(\delta^{aj_{2}...j_{q}}_{bi_{2}...i_{q}}\partial_{a}\partial_{b})^{-1}
\delta^{D-1}(\vec{x}-\vec{y}),\nonumber\\
\{C_{i_{1}...i_{q}}(\vec{x}),C_{j_{1}...j_{q}}(\vec{y})\}^{*}&=&\{D^{i_{1}...j_{p}}(\vec{x}),D^{j_{1}...j_{p}}(\vec{y})\}^{*}=0.
\end{eqnarray}

As in the previous case, we seek for the infinitesimal form of the
$SO(2)$ duality rotations acting on the canonical variables
$A_{\mu_{1}...\mu_{p}}$ and $C_{\mu_{1}...\mu_{p}}$ (rather than on
the "magnetic fields") and their conjugate momenta. When $p$ is odd
and $q$ is even or odd, the infinitesimal duality rotations  are given by
\begin{eqnarray}\label{rdi}
\delta E^{i_{1}...i_{p}}&=&\theta\frac{1}{q!}\varepsilon^{i_{1}...i_{p}jj_{1}...j_{q}}\partial_{j}C_{j_{1}...j_{q}},\nonumber\\
\delta
C_{i_{1}...i_{q}}&=&\theta\frac{1}{p!}\varepsilon^{i_{1}...i_{q}jj_{1}...j_{p}}\nabla^{-2}\partial_{i}E^{j_{1}...j_{p}},\nonumber\\
\delta D^{i_{1}...i_{q}}&=&\theta\frac{(-1)^{q+1}}{p!}\varepsilon^{i_{1}...i_{q}jj_{1}...j_{p}}\partial_{j}A_{j_{1}...j_{p}},\nonumber\\
\delta
A_{i_{1}...i_{p}}&=&\theta\frac{(-1)^{q}}{q!}\varepsilon^{i_{1}...i_{p}jj_{1}...j_{q}}\nabla^{-2}\partial_{j}D^{j_{1}...j_{q}}.
\end{eqnarray}
\\

The generator associated with these transformations is
\begin{eqnarray}\label{genomd}
G=\frac{1}{p!q!}\int
d^{D-1}x((-1)^{q}A_{i_{1}...i_{p}}\varepsilon^{i_{1}...i_{p}jj_{1}...j_{q}}\partial_{j}C_{j_{1}...j_{q}}\nonumber\\
+(-1)^{q+1}E^{i_{1}...i_{p}}\varepsilon^{i_{1}...i_{p}jj_{1}...j_{q}}\nabla^{-2}\partial_{j}D^{j_{1}...j_{q}}),
\end{eqnarray}

as can be verified.

As an example, let us consider
 the  generator for the "Maxwell-Klein-Gordon" theory in
$2+1$ dimensions, which corresponds to setting  $p=1$ and $q=0$ in equation (\ref{genomd}). In this case
the first order action is
\begin{eqnarray}
S=-\int
d^{3}x(E^{i}\dot{A}^{i}-D\dot{C}+\frac{1}{2}(E^{i}E^{i}+D^{2}+\varepsilon^{ijk}\partial_{j}A^{k}\varepsilon^{ilm}\partial_{j}A^{m}+\partial_{i}C\partial_{i}C))
\end{eqnarray}
where $C$ is the scalar field and $D$ its conjugate momenta. The
generator is then given by
\begin{eqnarray}
G=-\int d^{3}x(A^{i}\varepsilon^{ij}\partial_{j}C
+E^{i}\varepsilon^{ij}\nabla^{-2}\partial_{j}D).
\end{eqnarray}
Later, we shall take this simple case as a guide to interpret the geometrical content of the generator in an appropriate generalization of the
 Loop Representation that will be discussed.

To quantize the theory, we promote the canonical variables to
operators obeying  equal time canonical commutators obtained from the  Dirac brackets, and set up a Schroedinger equation with Hamiltonian given by

\begin{eqnarray}
\hat{H}=\int
d^{D-1}(\frac{(-1)^{p+1}}{2p!}\hat{E}^{i_{1}...i_{p}}\hat{E}^{i_{1}...i_{p}}+\frac{(-1)^{q+1}}{2q!}\hat{D}^{i_{1}...i_{q}}\hat{D}^{i_{1}...i_{q}}\\\nonumber
-\frac{(-1)^{p}}{2q!}\hat{B}^{i_{1}...i_{q}}\hat{B}^{i_{1}...i_{q}}-\frac{(-1)^{q}}{2p!}\hat{H}^{i_{1}...i_{p}}\hat{H}^{i_{1}...i_{p}}).
\end{eqnarray}
\\

\section{MASSIVE MODELS}
\subsection{Self Dual Massive Models}
In $D$ dimensions, the  theory of a massive $p$-form field is
described by the action
\begin{eqnarray}
S=\frac{(-1)^{p}}{2(p+1)!}\int
d^{D}xF_{\mu_{1}...\mu_{p+1}}F^{\mu_{1}...\mu_{p+1}}-(-1)^{p}\frac{m^{2}}{2p!}\int
d^{D}x A_{\mu_{1}...\mu_{p}}A^{\mu_{1}...\mu_{p}},
\end{eqnarray}
where $m$ is the mass  and
\begin{eqnarray}\label{fmnm}
F_{\mu\mu_{1}...\mu_{p}}=\frac{1}{p!}\partial_{[\mu}A_{\mu_{1}...\mu_{p}]},
\end{eqnarray}
with  $A_{\mu_{1}...\mu_{p}}$ being the antisymmetric field.

The equations of motion of this theory are given by
\begin{eqnarray}\label{emmm}
\partial_{\mu}F^{\mu\mu_{1}...\mu_{p}}=-m^{2}A^{\mu_{1}...\mu_{p}}.
\end{eqnarray}

When $D-1=2p$ the theory is self-dual, meaning that equation
(\ref{fmnm}) can be written as
\begin{eqnarray}\label{amnm}
\partial_{\mu}*A^{\mu\mu_{1}...\mu_{p}}=(-1)^{p}*F^{\mu_{1}...\mu_{p}},
\end{eqnarray}
which has the same form as equation (\ref{emmm}). Therefore, in the massive case, we find a duality symmetry  between
the potential $A$ and the Hodge dual of the field $F$ \cite{town},
unlike the massless case, where  duality relates
the electric and magnetic parts of the field $F$.

Expanding equations (\ref{emmm}) and (\ref{amnm}) we obtain
\begin{eqnarray}
\partial_{i}E^{ii_{1}...i_{p-1}}&=&-m^{2}A^{i_{1}...i_{p-1}0},\nonumber\\
(-1)^{p}\partial_{0}E^{i_{1}...i_{p}}+\frac{1}{(p+1)!}\varepsilon^{i_{1}...i_{p}ij_{1}...j_{p-1}}\partial_{i}B^{j_{1}...j_{p-1}}&=&-m^{2}A^{i_{1}...i_{p}},\nonumber\\
\frac{(-1)^{p+1}}{p!}\partial_{i}\varepsilon^{ii_{1}...i_{p-1}j_{1}...j_{p}}A^{j_{1}...j_{p}}&=&B^{i_{1}...i_{p-1}},\nonumber\\
\frac{(-1)^{p}}{p!}\varepsilon^{i_{1}...i_{p}j_{1}...j_{p}}\partial_{0}A^{j_{1}...j_{p}}-\frac{1}{(p+1)!}\varepsilon^{i_{1}...i_{p}ij_{1}...j_{p-1}}\partial_{i}A^{j_{1}...j_{p-1}0}&=&\frac{1}{p!}\varepsilon^{i_{1}...i_{p}j_{1}...j_{p}}E^{j_{1}...j_{p}},\nonumber\\
\end{eqnarray}
where $E^{i_{1}...i_{p}}=F^{i_{1}...i_{p}0}$ and $B^{i_{1}...i_{p-1}}=\frac{(-1)^{p}}{(p+1)!}\varepsilon^{i_{1}...i_{p-1}jj_{1}...j_{p}}F_{jj_{1}...j_{p}}$ are the generalized electric
and magnetic fields respectively.  As in the massless cases, it is found that the  equations of motion are invariant under duality transformations
 (this time between $F^{\mu_{1}...\mu_{p+1}}$ and $A^{\mu_{1}...\mu_{p}}$, rather than between electric and magnetic fields) belonging to
 the group $Z_{2}$ or $SO(2)$, depending on the evenness of the integer $p$. This fact can be summarized as follows. Let us introduce the objects
\[
\tilde{F}=\left(\begin{array}{c}
E^{i_{1}...i_{p}}\\
B^{i_{1}...i_{p-1}}
\end{array}
\right) \  \textrm{and} \   \tilde{A}=\left(\begin{array}{c}
m\varepsilon^{i_{1}...i_{p}j_{1}...j_{p}}A^{j_{1}...j_{p}}\\
-mA^{i_{1}...i_{p-1}0}
\end{array}
\right).
\]
Then, it can be seen that if $p$ is odd the equations of motion are invariant under $Z_{2}$ transformations given by
\begin{eqnarray}
\left(\begin{array}{c}
\tilde{F'}\\
\tilde{A'}\\
\end{array}\right)= \left(\begin{array}{cc}
0&1\\
1&0\\
\end{array}\right)
\left(\begin{array}{c}
F\\
A
\end{array}\right).
\end{eqnarray}
On the other hand, when $p$ is even, the invariance is under $SO(2)$ rotations
\begin{eqnarray}
\left(\begin{array}{c}
\tilde{F}\\
\tilde{A}
\end{array}\right)= \left(\begin{array}{cc}
cos\theta&sen\theta\\
-sen\theta&cos\theta\\
\end{array}\right)
\left(\begin{array}{c}
F\\
A
\end{array}\right).
\end{eqnarray}

As before, these results are in agreement with reference \cite{mene}. Carrying out the canonical
analysis we obtain the second class constraints
\begin{eqnarray}
\phi_{1}&=&E^{i_{1}...i_{p-1}0}=0,\nonumber\\
\phi_{2}&=&\partial_{i}E^{ii_{1}...i_{p-1}}+A^{i_{1}...i_{p-1}0}=0,
\end{eqnarray}
which should be familiar in the case $p=1$, that corresponds to
the Proca field. When $p$ is even, which is the case that exibits
duality as a Noether symmetry, the first order action results to
be
\begin{eqnarray}\label{accmass}
S=\int
d^{D-1}x(\frac{1}{p!}E^{i_{1}...i_{p}}\dot{A}_{i_{1}...i_{p}}-\frac{1}{2p!}E^{i_{1}...i_{p}}E^{i_{1}...i_{p}}-
\frac{1}{2(p-1)!}\partial_{i}E^{ii_{1}...i_{p-1}}\partial_{j}E^{ji_{1}...i_{p-1}}\nonumber\\
+\frac{1}{2(p+1)!}F_{i_{1}...i_{p+1}}F^{i_{1}...i_{p+1}}
-\frac{m^{2}}{2p!}A_{i_{1}...i_{p}}A^{i_{1}...i_{p}}).
\end{eqnarray}

The Dirac brackets between the canonical variables are
\begin{eqnarray}
\{A_{i_{1}...i_{p}}(\vec{x}),E^{i_{1}...i_{p}}(\vec{y})\}^{*}&=&\delta^{i_{1}...i_{p}}_{j_{1}...j_{p}}\delta^{D-1}(\vec{x}-\vec{y}),\nonumber\\
\{A_{i_{1}...i_{p}}(\vec{x}),A_{i_{1}...i_{p}}(\vec{y})\}^{*}&=&\{E^{i_{1}...i_{p}}(\vec{x}),E^{i_{1}...i_{p}}(\vec{y})\}^{*}=0,
\end{eqnarray}
as can be verified.

The infinitesimal $SO(2)$ duality
transformations of the  canonical variables can be written as
\begin{eqnarray}\label{afmad}
\delta E^{i_{1}...i_{p}}&=&\theta\frac{m}{p!}\varepsilon^{i_{1}...i_{p}j_{1}...j_{p}}A^{j_{1}...j_{p}},\nonumber\\
\delta A^{i_{1}...i_{p}}&=&-\theta\frac{1}{m
p!}\varepsilon^{i_{1}...i_{p}j_{1}...j_{p}}E^{i_{1}...i_{p}},
\end{eqnarray}
and the Noether  generator associated with these transformations  is
\begin{eqnarray}\label{genmad}
G=\int
d^{D-1}x(\frac{m}{2(p!)^{2}}\varepsilon^{i_{1}...i_{p}j_{1}...j_{p}}A_{i_{1}...i_{p}}A_{j_{1}...j_{p}}+
\frac{1}{2m(p!)^{2}}\varepsilon^{i_{1}...i_{p}j_{1}...j_{p}}E^{i_{1}...i_{p}}E^{j_{1}...j_{p}}).
\end{eqnarray}

Once again, in the quantum theory that corresponds to these models the equal time canonical commutators are obtained from the Dirac brackets, and the dynamics is governed by the Schroedinger equation
$i\frac{\partial|\Psi\rangle}{\partial t}=\hat{H}|\Psi\rangle$, with
Hamiltonian
\begin{eqnarray}
\hat{H}=\int
d^{D-1}x(\frac{1}{2p!}\hat{E}^{i_{1}...i_{p}}\hat{E}^{i_{1}...i_{p}}+\frac{1}{2(p-1)!}\partial_{i}\hat{E}^{ii_{1}...i_{p-1}}\partial_{j}\hat{E}^{ji_{1}...i_{p-1}}
\nonumber\\
+\frac{1}{2(p-1)!}\hat{B}^{i_{1}...i_{p-1}}\hat{B}^{i_{1}...i_{p-1}}+\frac{m^{2}}{2p!}\hat{A}_{i_{1}...i_{p}}\hat{A}^{i_{1}...i_{p}}).
\end{eqnarray}
The lowest dimensionality example of these models corresponds to
the simple harmonic oscillator. In that case, $D=1$ (the temporal
dimension) and $p=0$. The first order action can be obtained by
replacing $E$ by $\frac{p}{\sqrt{m_{osc}}}$, $A$ by
$\sqrt{m_{osc}}x$ and $m$ by $\sqrt{{\frac{k}{m_{osc}}}}$, which
yields
\begin{eqnarray}
S=\int dt (p\dot{x} -\frac{p^{2}}{2m_{osc}}-\frac{1}{2}kx^{2}),
\end{eqnarray}
where $p$ and $x$ are the canonical variables, $m_{osc}$ is the oscillator
mass and $k$ is the spring constant. Therefore, the  generator of duality is
given by
\begin{eqnarray}
G=\frac{1}{2}\sqrt{km_{osc}}x^{2}+\frac{1}{2\sqrt{km_{osc}}}p^{2}.
\end{eqnarray}

\subsection{Dual Massive
Models} Finally, we consider theories comprising a pair of uncoupled massive
$p$ and $q$-forms, that map one into the other under the action of
duality transformations. We take the action as
\begin{eqnarray}
S=\frac{(-1)^{p}}{2(p+1)!}\int
d^{D}xF_{\mu_{1}...\mu_{p+1}}F^{\mu_{1}...\mu_{p+1}}-(-1)^{p}\frac{m^{2}}{2p!}\int
d^{D}x A_{\mu_{1}...\mu_{p}}A^{\mu_{1}...\mu_{p}}\nonumber\\
+\frac{(-1)^{q}}{2(q+1)!}\int
d^{D}xG_{\mu_{1}...\mu_{q+1}}G^{\mu_{1}...\mu_{q+1}}-(-1)^{q}\frac{m^{2}}{2q!}\int
d^{D}x C _{\mu_{1}...\mu_{q}}C^{\mu_{1}...\mu_{q}},
\end{eqnarray}
where $m$ is the mass. The field strengths are
\begin{eqnarray}\label{fmnmd}
F_{\mu\mu_{1}...\mu_{p}}=\frac{1}{p!}\partial_{[\mu}A_{\mu_{1}...\mu_{p}]},\nonumber\\
G_{\mu\mu_{1}...\mu_{q}}=\frac{1}{q!}\partial_{[\mu}C_{\mu_{1}...\mu_{q}]},
\end{eqnarray}
while  $A_{\mu_{1}...\mu_{p}}$ and $C_{\mu_{1}...\mu_{q}}$ correspond to the antisymmetric potentials.
\\

This time the equations of motion are given by
\begin{eqnarray}
\partial_{\mu}F^{\mu\mu_{1}...\mu_{p}}=-m^{2}A^{\mu_{1}...\mu_{p}}\nonumber\\
\partial_{\mu}G^{\mu\mu_{1}...\mu_{p}}=-m^{2}C^{\mu_{1}...\mu_{q}}\label{emmmd}.
\end{eqnarray}

Taken into account that equations (\ref{fmnmd}) may be written as
\begin{eqnarray}\label{amnmd}
\partial_{\mu}*A^{\mu\mu_{1}...\mu_{p}}=(-1)^{p}*F^{\mu_{1}...\mu_{p}}\nonumber\\
\partial_{\mu}*C^{\mu\mu_{1}...\mu_{q}}=(-1)^{q}*F^{\mu_{1}...\mu_{q}},
\end{eqnarray}
we see that for $D-1=p+q$ there is a duality symmetry between the $p$ and $q$ fields of the theory. As in the previous cases, this symmetry
can be studied at the level of the canonical variables by expanding (\ref{emmmd}) and (\ref{amnmd}) to obtain
\begin{eqnarray}
\partial_{i}E^{ii_{1}...i_{p-1}}&=&-m^{2}A^{i_{1}...i_{p-1}0},\nonumber\\
(-1)^{p}\partial_{0}E^{i_{1}...i_{p}}+\frac{1}{(q-1)!}\varepsilon^{i_{1}...i_{p}ij_{1}...j_{q-1}}\partial_{i}B^{j_{1}...j_{q-1}}&=&-m^{2}A^{i_{1}...i_{p}},\nonumber\\
\frac{(-1)^{p+1}}{p!}\partial_{i}\varepsilon^{ii_{1}...i_{q-1}j_{1}...j_{p}}A^{j_{1}...j_{p}}&=&B^{i_{1}...i_{q-1}},\nonumber\\
\frac{(-1)^{p}}{p!}\varepsilon^{i_{1}...i_{q}j_{1}...j_{p}}\partial_{0}A^{j_{1}...j_{p}}-\frac{1}{(p-1)!}\varepsilon^{i_{1}...i_{q}ij_{1}...j_{p-1}}\partial_{i}A^{j_{1}...j_{p-1}0}&=&\frac{1}{p!}\varepsilon^{i_{1}...i_{q}j_{1}...j_{p}}E^{j_{1}...j_{p}},\nonumber\\
\partial_{i}D^{ii_{1}...i_{q-1}}&=&-m^{2}A^{i_{1}...i_{q-1}0},\nonumber\\
(-1)^{q}\partial_{0}D^{i_{1}...i_{q}}+\frac{1}{(q+1)!}\varepsilon^{i_{1}...i_{q}ij_{1}...j_{q-1}}\partial_{i}H^{j_{1}...j_{q-1}}&=&-m^{2}A^{i_{1}...i_{q}},\nonumber\\
\frac{(-1)^{q+1}}{q!}\partial_{i}\varepsilon^{ii_{1}...i_{p-1}j_{1}...j_{q}}C^{j_{1}...j_{q}}&=&H^{i_{1}...i_{p-1}},\nonumber\\
\frac{(-1)^{q}}{q!}\varepsilon^{i_{1}...i_{p}j_{1}...j_{q}}\partial_{0}C^{j_{1}...j_{q}}-\frac{1}{(q-1)!}\varepsilon^{i_{1}...i_{p}ij_{1}...j_{q-1}}\partial_{i}C^{j_{1}...j_{q-1}0}&=&\frac{1}{q!}\varepsilon^{i_{1}...i_{p}j_{1}...j_{q}}D^{j_{1}...j_{q}}.\nonumber\\
\end{eqnarray}

Here, $E^{i_{1}...i_{p}}=F^{i_{1}...i_{p}0}$ and $D^{i_{1}...i_{q}}=G^{i_{1}...i_{q}0}$ are the generalized electric
fields, while\\
$B^{i_{1}...i_{q-1}}=\frac{(-1)^{p}}{(p+1)!}\varepsilon^{i_{1}...i_{q-1}jj_{1}...j_{p}}F_{jj_{1}...j_{p}}$ and
$H^{i_{1}...i_{p-1}}=\frac{(-1)^{q}}{(q+1)!}\varepsilon^{i_{1}...i_{p-1}jj_{1}...j_{q}}G_{jj_{1}...j_{q}}$
are the magnetic ones.

Following already familiar steps, we define the objects
\[
\tilde{F}=\left(\begin{array}{c}
E^{i_{1}...i_{p}}\\
B^{i_{1}...i_{q-1}}\\
D^{i_{1}...i_{q-1}}\\
H^{i_{1}...i_{p}}
\end{array}
\right) \  \textrm{and} \   \tilde{A}=\left(\begin{array}{c}
m\varepsilon^{i_{1}...i_{p}j_{1}...j_{q}}C^{j_{1}...j_{q}}\\
-mC^{i_{1}...i_{q-1}0}\\
m\varepsilon^{i_{1}...i_{q}j_{1}...j_{p}}A^{j_{1}...j_{p}}\\
-mA^{i_{1}...i_{p-1}0}
\end{array}
\right).
\]
Then, for $p$ odd and $q$ even or odd  the equations of motion are
invariant under $Z_{2}$ transformations
\begin{eqnarray}
\left(\begin{array}{c}
\tilde{F'}\\
\tilde{A'}\\
\end{array}\right)= \left(\begin{array}{cc}
0&1\\
1&0\\
\end{array}\right)
\left(\begin{array}{c}
F\\
A
\end{array}\right),
\end{eqnarray}

while when $p$ and $q$ are even, the  invariance is  under $SO(2)$ duality rotations
\begin{eqnarray}
\left(\begin{array}{c}
\tilde{F}\\
\tilde{A}
\end{array}\right)= \left(\begin{array}{cc}
cos\theta&sen\theta\\
-sen\theta&cos\theta\\
\end{array}\right)
\left(\begin{array}{c}
F\\
A
\end{array}\right).
\end{eqnarray}

The canonical formulation of Dirac yields the second class constraints
\begin{eqnarray}
\phi_{1}&=&E^{i_{1}...i_{p-1}0}=0,\nonumber\\
\phi_{2}&=&D^{i_{1}...i_{q-1}0}=0,\nonumber\\
\phi_{3}&=&\partial_{i}E^{ii_{1}...i_{p-1}}+A^{i_{1}...i_{p-1}0}=0,\nonumber\\
\phi_{4}&=&\partial_{i}D^{ii_{1}...i_{q-1}}+C^{i_{1}...i_{q-1}0}=0,
\end{eqnarray}
 When $p$ and $q$ are even, which is the interesting case from the point of view of the Noether theorem, the first order action is given by
\begin{eqnarray}
S&=&\int
d^{D-1}x(\frac{1}{p!}E^{i_{1}...i_{p}}\dot{A}_{i_{1}...i_{p}}-\frac{1}{2p!}E^{i_{1}...i_{p}}E^{i_{1}...i_{p}}-
\frac{1}{2(p-1)!}\partial_{i}E^{ii_{1}...i_{p-1}}\partial_{j}E^{ji_{1}...i_{p-1}}\nonumber\\
&+&\frac{1}{2(p+1)!}F_{i_{1}...i_{p+1}}F^{i_{1}...i_{p+1}}
-\frac{m^{2}}{2p!}A_{i_{1}...i_{p}}A^{i_{1}...i_{p}}+\frac{1}{q!}D^{i_{1}...i_{q}}\dot{C}_{i_{1}...i_{q}}-\frac{1}{2q!}D^{i_{1}...i_{q}}D^{i_{1}...i_{q}}\nonumber\\
&-&
\frac{1}{2(q-1)!}\partial_{i}D^{ii_{1}...i_{q-1}}\partial_{j}D^{ji_{1}...i_{q-1}}+\frac{1}{2(q+1)!}G_{i_{1}...i_{q+1}}G^{i_{1}...i_{q+1}}\nonumber\\
&-&\frac{m^{2}}{2q!}C_{i_{1}...i_{q}}C^{i_{1}...i_{q}}),
\end{eqnarray}
and the fundamental Dirac brackets are
\begin{eqnarray}
\{A_{i_{1}...i_{p}}(\vec{x}),E^{j_{1}...j_{p}}(\vec{y})\}^{*}&=&\delta^{i_{1}...i_{p}}_{j_{1}...j_{p}}\delta^{D-1}(\vec{x}-\vec{y}),\nonumber\\
\{C_{i_{1}...i_{q}}(\vec{x}),D^{j_{1}...j_{q}}(\vec{y})\}^{*}&=&\delta^{i_{1}...i_{p}}_{j_{1}...j_{p}}\delta^{D-1}(\vec{x}-\vec{y}),\nonumber\\
\{A_{i_{1}...i_{p}}(\vec{x}),A_{j_{1}...j_{p}}(\vec{y})\}^{*}&=&\{E^{i_{1}...i_{p}}(\vec{x}),E^{j_{1}...j_{p}}(\vec{y})\}^{*}=0,\nonumber\\
\{C_{i_{1}...i_{q}}(\vec{x}),C_{j_{1}...j_{q}}(\vec{y})\}^{*}&=&\{D^{i_{1}...i_{q}}(\vec{x}),D^{j_{1}...j_{q}}(\vec{y})\}^{*}=0.
\end{eqnarray}

The behavior of the canonical variables under infinitesimal duality
rotations can be written as
\begin{eqnarray}\label{agma}
\delta E^{i_{1}...i_{p}}=\theta\frac{m}{q!}\varepsilon^{i_{1}...i_{p}j_{1}...j_{q}}C_{j_{1}...j_{q}},\nonumber\\
\delta C^{i_{1}...i_{q}}=-\theta\frac{1}{m
p!}\varepsilon^{i_{1}...i_{q}j_{1}...j_{p}}E^{j_{1}...j_{p}},\nonumber\\
\delta D^{i_{1}...i_{q}}=\theta\frac{m}{p!}\varepsilon^{i_{1}...i_{q}j_{1}...j_{p}}A_{j_{1}...j_{p}},\nonumber\\
\delta A^{i_{1}...i_{p}}=-\theta\frac{1}{m
q!}\varepsilon^{i_{1}...i_{p}j_{1}...j_{q}}D_{j_{1}...j_{q}}.
\end{eqnarray}

The Noether charge that generates these transformations is
\begin{eqnarray}\label{genmd}
G=\int
d^{D-1}x(\frac{m}{p!q!}\varepsilon^{i_{1}...i_{p}j_{1}...j_{q}}A_{i_{1}...i_{p}}C_{j_{1}...j_{q}}+\frac{1}{mp!q!}\varepsilon^{i_{1}...i_{p}j_{1}...j_{q}}E^{i_{1}...i_{p}}D^{j_{1}...j_{q}}).
\end{eqnarray}

The quantum theory is obtained as in the previous cases. The equal time canonical commutators are based on the Dirac brackets and the dynamics is given by the Schroedinger equation with Hamiltonian

\begin{eqnarray}
\hat{H}&=&\int
d^{D-1}x(\frac{1}{2p!}\hat{E}^{i_{1}...i_{p}}\hat{E}^{i_{1}...i_{p}}+\frac{1}{2(p-1)!}\partial_{i}\hat{E}^{ii_{1}...i_{p-1}}\partial_{j}\hat{E}^{ji_{1}...i_{p-1}}
+\frac{1}{2(q-1)!}\hat{B}^{i_{1}...i_{q-1}}\hat{B}^{i_{1}...i_{q-1}}\nonumber\\
&-&\frac{m^{2}}{2p!}\hat{A}_{i_{1}...i_{p}}\hat{A}^{i_{1}...i_{p}}+\frac{1}{2q!}\hat{D}^{i_{1}...i_{q}}\hat{D}^{i_{1}...i_{q}}+\frac{1}{2(q-1)!}\partial_{i}\hat{D}^{ii_{1}...i_{q-1}}\partial_{j}\hat{D}^{ji_{1}...i_{q-1}}\nonumber\\
&+&\frac{1}{2(p-1)!}\hat{H}^{i_{1}...i_{p-1}}\hat{H}^{i_{1}...i_{p-1}}-\frac{m^{2}}{2q!}\hat{C}_{i_{1}...i_{q}}\hat{C}^{i_{1}...i_{q}}).
\end{eqnarray}

This completes our review of duality transformations in Abelian
theories. It should be stressed that all this construction is based
on the study of the $D=4$ Maxwell case of reference \cite{deser}.
The different aspects of the duality transformations, depending on
whether the theory is massive or not, on the number of different
fields appearing, and on the order of the fields and the dimension
of the space time, can also be found in previous works
\cite{wotza,nor,mene}. However,  as far as we know,
 the explicit form of the generator $G$ for all these cases, and the detailed canonical analysis of the different models,
 have not been reported previously.

\section{PATH REPRESENTATION AND GENERALIZATIONS}
In this section we discuss the path-space representation and certain
generalizations of it adapted to deal with both the massless and massive models discussed in the previous sections. The purpose of this study is to realize the generators of duality transformations of these models in a geometric representation, in order to get some insight into their geometrical meaning.

We begin by recalling that the Abelian path space can be described
as the set of certain equivalence classes of curves $\gamma$ in a
manifold, which we take as $R^{n}$ \cite{dibargambtri, gambpull,
gambtri, alelore}. The equivalence relation is given in terms of the
so called form factor $T^{i}(\vec{x},\gamma)$ of the curves
\begin{eqnarray}
T^{i}(\vec{x},\gamma)=\int_{\gamma}dy^{i}\delta^{n}(\vec{x}-\vec{y})
\end{eqnarray}
as follows: $\gamma$ and $\gamma'$ are said to be equivalent (i.e.,
they represent the same path) if their form factors coincide. Closed
curves give raise to a subspace of the path space: the loop space.
It can be seen that usual composition of curves translates into a
composition of paths that endows path space with an Abelian group
structure \cite{gambpull}.

The path representation arises when one realizes the canonical field
operators onto  path-dependent wave functionals $\Psi[\gamma]$. We
define the path and loop derivatives $\delta_{i}(\vec{x})$ and
$\triangle_{ij}(\vec{x})$ by \cite{gambpull}
\begin{eqnarray}
u^{i}\delta_{i}(\vec{x})\Psi\left[\gamma\right]\equiv\Psi\left[\gamma\circ
u_{\vec{x}}\right]-\Psi\left[\gamma\right],\nonumber\\
\frac{1}{2}\sigma^{ij}(\vec{x})\triangle_{ij}(\vec{x})\Psi\left[\gamma\right]=\Psi\left[\gamma\circ\delta
c\right]-\Psi\left[\gamma\right],
\end{eqnarray}
where $\circ$ denotes the path space product. The derivative
$\delta_{i}(\vec{x})$ ($\triangle_{ij}(\vec{x})$) measures the
change in the path-dependent wave functional when an infinitesimal
path $\delta u$ (infinitesimal loop $\delta c$) is attached to its
argument $\gamma$ at the point $\vec{x}$. It is understood that
these changes are considered up to first order in the infinitesimal
vector $\vec{u}$  associated with the small path, or in the surface
element  $\sigma^{ij}=u^{i}v^{j}-u^{j}v^{i}$ (depending on the case)
 generated by the infinitesimal vectors $\vec{u}$ and $\vec{v}$ that
define the small loop $\delta c$. It can be shown that both
derivatives are related by
\begin{eqnarray}
\partial_{i}\delta_{j}(\vec{x})-\partial_{j}\delta_{i}(\vec{x})=\triangle_{ij}(\vec{x}).
\end{eqnarray}
\\

As an example of how these operators work, we calculate the path derivative of the form
factor. One has
\begin{eqnarray}
T^{i}(\vec{x},\gamma\circ u_{\vec{y}})&=&\int_{\gamma\circ
u_{\vec{y}}}d z^{i}\delta^{n}(\vec{x}-\vec{y})\nonumber\\
&=&T^{i}(\vec{x},\gamma)+u^{j}\delta^{i}_{j}\delta^{n}(\vec{x}-\vec{y}).
\end{eqnarray}
Hence
\begin{eqnarray}
\delta_{i}(\vec{y})T^{j}(\vec{x},\gamma)=\delta^{i}_{j}\delta^{n}(\vec{x}-\vec{y}).
\end{eqnarray}

The Abelian path-space construction may be extended to the case of
$p$-surfaces ($p>1$) (see references \cite{alelore} for the $p=2$
generalization). In these cases the geometric space can be described
as a set of equivalence classes of $p$-surfaces $\Sigma$ labeled by
the $p$-surface  form-factor, which is defined as

\begin{eqnarray}
T^{i_{1}...i_{p}}(\vec{x},\Sigma)=\int_{\Sigma}d\Sigma^{i_{1}...i_{p}}_{\vec{y}}\delta^{D-1}(\vec{x}-\vec{y}).
\end{eqnarray}
Here, $d\Sigma^{i_{1}...i_{p}}$ is the surface element of
$\Sigma$. The surfaces $\Sigma$ and $\Sigma'$ will be considered
equivalent if their form factor are the same. As in the $p=1$
case, closed $p$-surfaces give rise to a subspace of the $p$
-surface space: the closed $p$-surface space.

Now we can define the open $p$-surface derivative
$\delta_{i_{1}...i_{p}}(\vec{x})$
\begin{eqnarray}
\sigma^{i_{1}...i_{p}}\delta_{i_{1}...i_{p}}(\vec{x})\Psi\left[\Sigma\right]\equiv\Psi\left[\Sigma\circ
\sigma_{\vec{x}}\right]-\Psi\left[\Sigma\right],\\
\end{eqnarray}

that measures the change of the path-dependent functional
$\Psi\left[\Sigma\right]$ when an infinitesimal surface of area
$\sigma^{i_{1}...i_{p}}$  is appended  to its argument $\Sigma$ at
the point $\vec{x}$. As in the case of the path derivative, it is
understood that these changes are considered up to first order in
$\sigma^{i_{1}...i_{p}}$.

Also, we define the closed $p$-surface derivative,
$\triangle_{i_{1}...i_{p+1}}(\vec{x})$, which instead of an open
surface attaches an infinitesimal closed one to the argument of the
functional. It is given by
\begin{eqnarray}
\sigma^{i_{1}...i_{p+1}}\triangle_{i_{1}...i_{p+1}}(\vec{x})\Psi\left[\Sigma\right]\equiv\Psi
\left[\Sigma\circ\delta \sigma\right]-\Psi\left[\Sigma\right].
\end{eqnarray}

It can be seen that both derivatives are related by
\begin{eqnarray}
\triangle_{ii_{1}...i_{p}}(\vec{x})=\frac{1}{p!}\partial_{[i}\delta_{i_{1}...i_{p}]}(\vec{x}).
\end{eqnarray}

The open $p$-surface derivative of the generalized form
factor is then given by
\begin{eqnarray}
\delta_{j_{1}...j_{p}}(\vec{y})T^{i_{1}...i_{p}}(\vec{x},\Sigma)=\delta^{i_{1}...i_{p}}_{j_{1}...j_{p}}\delta^{D-1}(\vec{x}-\vec{y}).
\end{eqnarray}
It should be clear that this construction is nothing but a
generalization of the ideas underlying the path-space formulation to
the case of extended objects of higher dimensions. Later, we shall
briefly refer to an extension of this framework in the opposite
sense: the "signed points" space \cite{leal}, in which the
geometrical objects of interest are collections of points rather
than paths or surfaces.

With these tools at hand we are ready to represent the operators acting on $p$-surface dependent wave functionals $\Psi\left[\Sigma\right]$.
For the self-dual massless models we take

\begin{eqnarray}
\hat{E}^{i_{1}...i_{p}}(\vec{x})=T^{i_{1}...i_{p}}(\vec{x},\Sigma)\nonumber\\
\hat{B}^{i_{1}...i_{p}}(\vec{x})=-\frac{i}{(p+1)!}\varepsilon^{i_{1}...i_{p}jj_{1}...j_{p}}\triangle_{jj_{1}...j_{p}}(\vec{x}).
\end{eqnarray}
It can be shown that this prescription is a representation of the
algebra of the basic observables $\hat{B}^{i_{1}...i_{p}}(\vec{x})$
and $\hat{E}^{j_{1}...j_{p}}(\vec{y})$ that arises from equation
(\ref{algae}), namely
\begin{eqnarray}
\left[\hat{B}^{i_{1}...i_{p}}(\vec{x}),\hat{E}^{j_{1}...j_{p}}(\vec{y})\right]=-i\varepsilon^{i_{1}...i_{p}jj_{1}...j_{p}}\partial^{x}
_{j}\delta^{D-1}(\vec{x}-\vec{y}).
\end{eqnarray}

It is well known that in the Maxwell case without sources  the
Gauss Law constraint $\partial_{i}E^{i}=0$ reduces path-space to
the loop-space \cite{dibargambtri}. This is so because in the
absence of sources, paths, which are quantum Faradays lines, have
no charge to emanate from or to arrive, and therefore must be
closed. In the generalized case, since $E^{i_{1}...i_{p}}$ is also
transverse, the $p$-surface space must be also reduced to the
subspace of closed $p$-surfaces. On the other hand, all the
observables can be written down in terms of the gauge invariant
basic observables, which are the (generalized) electric and
magnetic fields. This is so even for the duality generator (which
is also gauge invariant), which in the geometric representation of
closed $p$-surfaces is given by

\begin{eqnarray}
\hat{G}=-\frac{1}{2\Omega_{s}p!(p+1)!}\int d^{D-1}x\int
d^{D-1}y\varepsilon^{i_{1}...i_{p}k_{1}...k_{p+1}}\triangle_{k_{1}...k_{p+1}}(\vec{x})
\triangle_{j_{1}...j_{p}j}(\vec{y})\frac{(x-y)^{j}}{|\vec{x}-\vec{y}|^{D-1}}+\nonumber\\
+\frac{D-3}{2\Omega_{s}(p!)^{2}}\oint_{\Sigma}
d\Sigma^{i_{1}...i_{p}}_{\vec{x}}\oint_{\Sigma}
d\Sigma^{j_{1}...j_{p}}_{\vec{y}}\varepsilon^{i_{1}...i_{p}j_{1}...j_{p}j}\frac{(x-y)^{j}}{|\vec{x}-\vec{y}|^{D-1}}.
\end{eqnarray}
Here, $\Omega_{s}$ is the volume of the $D-1$-dimensional unitary
sphere. The second term is the generalized self-linking number
\cite{rolf} of the closed $p$ surface $\Sigma^{i_{1}...i_{p}}$
(recall that $D=2p+2$). In particular, for $p=1$ and $D=4$ we obtain
the realization of the duality generator for Maxwell theory  in the
loop representation
\begin{eqnarray}
\hat{G}=-\frac{1}{16\pi}\int d^{3}x\int d^{3}y\varepsilon^{ijk}\triangle_{jk}(\vec{x})\triangle_{il}(\vec{y})\frac{(x-y)^{l}}{|\vec{x}-\vec{y}|^{3}}\nonumber\\
-\frac{1}{8\pi}\oint_{\gamma} dx^{i}\oint_{\gamma}
dy^{j}\varepsilon^{ijk}\frac{(x-y)^{k}}{|\vec{x}-\vec{y}|^{3}}.
\end{eqnarray}

The second term, which is the loop self-linking number
\cite{rolf}, has an intuitive geometrical interpretation. It
measures the oriented number of times that the loop $\gamma_{i}$
intercepts any surface bordered by itself. The interpretation of
this invariant in the general case is similar: it counts the
oriented number of times that the $p$-surface $\Sigma$ cuts any
$p+1$-surface whose boundary is $\Sigma$ itself.

We should recall that the existence of this generator holds for
$p$ odd. For $p$ even, there is also a generalized loop
representation of the massless self-dual models, however, since
duality transformations are discrete in that case, there is no
room for an infinitesimal  generator.

Now we turn our attention to dual massless models. In these cases,
the geometric representation must be built in terms of wave
functionals
 $\Psi\left[\Sigma_{p},\Sigma_{q}\right]$ that depend on two  surfaces $\Sigma_{p},\Sigma_{q}$ of different dimension. Now we have to kinds of surface operators, each acting on
one of the arguments ($\Sigma_{p}$ or $\Sigma_{q}$) of the wave
functional. Then, following similar steps
 as before, we
prescribe the realization
\begin{eqnarray}
\hat{E}^{i_{1}...i_{p}}(\vec{x})=T^{i_{1}...i_{p}}(\vec{x},\Sigma_{p}),\nonumber\\
\hat{B}^{i_{1}...i_{q}}(\vec{x})=i\frac{(-1)^{p}}{(p+1)!}\varepsilon^{i_{1}...i_{q}jj_{1}...j_{p}}\triangle_{jj_{1}...j_{p}}(\vec{x}),\\
\hat{D}^{i_{1}...i_{q}}(\vec{x})=T^{i_{1}...i_{q}}(\vec{x},\Sigma_{q}),\nonumber\\
\hat{H}^{i_{1}...i_{p}}(\vec{x})=i\frac{(-1)^{q}}{(q+1)!}\varepsilon^{i_{1}...i_{p}jj_{1}...j_{q}}\triangle_{jj_{1}...j_{q}}(\vec{x}),
\end{eqnarray}
which fulfils the algebra of elementary observables
(i.e., generalized electric and magnetic fields)
obtained from the canonical algebra
\begin{eqnarray}
\left[\hat{B}^{i_{1}...i_{q}}(\vec{x}),\hat{E}^{j_{1}...j_{p}}(\vec{y})\right]=-i\varepsilon^{i_{1}...i_{q}jj_{1}...j_{p}}\partial^{x}
_{j}\delta^{D-1}(\vec{x}-\vec{y}),\nonumber\\
\left[\hat{H}^{i_{1}...i_{p}}(\vec{x}),\hat{D}^{j_{1}...j_{q}}(\vec{y})\right]=-i\varepsilon^{i_{1}...i_{p}jj_{1}...j_{q}}\partial^{x}
_{j}\delta^{D-1}(\vec{x}-\vec{y}).
\end{eqnarray}
 As before, the $p$-surfaces and $q$-surfaces spaces
result to be reduced to the subspaces of closed surfaces, because
both $E^{i_{1}...i_{p}}$ and $D^{i_{1}...i_{q}}$ are transverse in
virtue of the generalized Gauss laws that they obey
($\partial_{i_{1}}E^{i_{1}...i_{p}}=
\partial_{i_{1}}D^{i_{1}...i_{q}}=0$. On the other hand, the duality generator, in the cases were
the duality group is $SO(2)$ (i.e., when $p$ is odd and $q$ is
even or odd) is given by
\begin{eqnarray}
\hat{G}=\frac{(-1)^{q}}{\Omega_{s}(p+1)!q!}\int d^{D-1}x\int
d^{D-1}y\varepsilon^{jj_{1}...j_{p}k_{1}...k_{q}}\triangle_{jj_{1}...j_{p}}(\vec{y})\triangle_{k_{1}...k_{q}k}(\vec{x})\frac{(x-y)^{k}}{|\vec{x}-\vec{y}|^{D-1}}
\nonumber\\
+\frac{D-3}{\Omega_{s}p!q!}\oint_{\Sigma}
d\Sigma^{i_{1}...i_{p}}_{\vec{x}}\oint_{\Sigma}
d\Sigma^{j_{1}...j_{q}}_{\vec{y}}\varepsilon^{i_{1}...i_{p}j_{1}...j_{q}j}\frac{(x-y)^{j}}{|\vec{x}-\vec{y}|^{D-1}}.
\end{eqnarray}
Again, the second term admits a simple geometrical interpretation.
It corresponds to  the generalized linking number between the
surfaces $\Sigma_{p}$ and $\Sigma_{q}$ \cite{rolf}, that counts
the oriented number of times that  the $p$-surface  and a region
enclosed by the $q$-surface intercept in the $p+q$ dimensional
space.

Unlike the massless cases, where the basic operators to be realized
are the generalized electric and magnetic fields, in the massive
ones we have to realize the potentials, together with the electric
fields, which are the canonical variables. For the self-dual massive
case we set
\begin{eqnarray}
\hat{E}^{i_{1}...i_{p}}(\vec{x})=T^{i_{1}...i_{p}}(\vec{x},\Sigma)\nonumber\\
\hat{A}_{i_{1}...i_{p}}(\vec{x})=i\delta_{i_{1}...i_{p}}(\vec{x}),
\end{eqnarray}
which satisfy the canonical algebra. In contrast with
the massless models, the surface-space involved is not the subspace
of closed surfaces, since this time there is no Gauss Law to force
the "electric fields" to be transverse. As before, we are specially
interested in the generator of duality rotations, which results to
be
\begin{eqnarray}
\hat{G}=-\frac{m}{2(p!)^{2}}\int d^{D-1}x\varepsilon^{i_{1}...i_{p}j_{1}...j_{p}}\delta(\vec{x})_{i_{1}...i_{p}}\delta(\vec{x})_{j_{1}...j_{p}}\nonumber\\
+\frac{\varepsilon^{i_{1}...i_{p}j_{1}...j_{p}}}{2m(p!)^{2}}\int
d\Sigma_{\vec{x}}^{i_{1}...i_{p}}\int
d\Sigma_{\vec{y}}^{j_{1}...j_{p}}\delta^{D-1}(x-y).
\end{eqnarray}

 This time the second term is not a linking number. Instead, it is the self-intersection number of the closed
 surface $\Sigma_{p}$ \cite{rolf}
 which counts the oriented number of times that, in $D-1=2p$ dimensions, a $p$-surface punctures
  itself. It should be noticed that for
 $p$ odd, the "would be duality-generator" vanishes, in agreement with the fact that in that case duality is a
 discrete symmetry. For instance, for $p=1$,
 which corresponds to curves, it can be readily seen that this self-intersection number vanishes.

Finally, let us discuss the dual massive cases. Now we set
\begin{eqnarray}
\hat{E}^{i_{1}...i_{p}}(\vec{x})&=&T^{i_{1}...i_{p}}(\vec{x},\Sigma_{p}),\nonumber\\
\hat{D}^{i_{1}...i_{q}}(\vec{x})&=&T^{i_{1}...i_{q}}(\vec{x},\Sigma_{q}),\nonumber\\
\hat{A}_{i_{1}...i_{p}}(\vec{x})&=&i\delta_{i_{1}...i_{p}}(\vec{x}),\nonumber\\
\hat{C}_{i_{1}...i_{q}}(\vec{x})&=&i\delta_{i_{1}...i_{q}}(\vec{x}).
\end{eqnarray}
which realizes the canonical algebra. These operators act onto
wave functionals that depend on two different open surfaces
$\Sigma_{p}$ and $\Sigma_{q}$ of dimensions $p$ and $q$
respectively. The generator of duality transformations is given by
\begin{eqnarray}
\hat{G}=-\frac{m}{p!q!}\int d^{D-1}x\varepsilon^{i_{1}...i_{p}j_{1}...j_{q}}\delta_{i_{1}...i_{p}}(\vec{x})\delta_{j_{1}...j_{q}}(\vec{x})\nonumber\\
+\frac{\varepsilon^{i_{1}...i_{p}j_{1}...j_{q}}}{mp!q!}\int_{\Sigma}
d\Sigma_{\vec{x}}^{i_{1}...i_{p}}\int_{\Sigma'}
d\Sigma_{\vec{x}'}^{j_{1}...j_{q}}\delta^{D-1}(\vec{x}-\vec{y}).
\end{eqnarray}

This time, the second term measures the number of oriented
intersections between the  surfaces $\Sigma^{i_{1}...i_{p}}$ and
$\Sigma^{j_{1}...j_{q}}$.
\\

To conclude, let us discuss briefly how to deal with the "loop
representation" for $0$-forms.  When $p=0$, the canonical operators
can be realized in the space of signed points introduced in
reference \cite{leal} . There, both the Abelian and non-Abelian
cases were considered. For our purposes, we shall just take some
results that we need to deal with the Abelian situation. We consider
the set  whose elements are unordered lists $X$ of "signed" points
$x^{s}$, where
 $s=\pm$. A list would be, for instance: $X=x_{1}^{+}, x_{2}^{-},...,x_{s}^{+}$. The points can be though as boundaries of oriented paths
  (maybe starting or ending at the spatial infinity), inasmuch as loops can be seen as boundaries of $2$- surfaces. The form factor of a
  list is defined as
\begin{eqnarray}
T(x,X)=\sum_{a}s_{a}\delta^{n}(x-x_{a}),
\end{eqnarray}
where $s_{a}$ is the sign of the point at position $x_{a}$. Then we declare that two
lists of points $X$ and $Y$ are equivalent if
\begin{eqnarray}
T(x,X)=T(x,Y).
\end{eqnarray}

Now we proceed as before. Considering functionals $\Psi[X]$ that depend on lists, we can define the
object $a(x^{s})$ which act on $\Psi[X]$ adding a signed point in
$X$. In \cite{leal} it is shown that we can also define an operator that
measures the change of $\Psi[X]$ when  an
infinitesimal list of points $\delta Y$ is added, by means of
\begin{eqnarray}
\triangle_{i}(x)\equiv a(x)\frac{\partial}{\partial x^{i}}a(x)^{-1}.
\end{eqnarray}
This object, the "dipole derivative",   is analogous to the loop
derivative in the loop representation.

As an example, we use this representation to realize the duality
generator for the "Maxwell-Klein-Gordon" theory in $D=3$, which
was discussed at the end of section $2.2$. The operators can be
realized as
\begin{eqnarray}
\hat{E^{i}}(\vec{x})=T^{i}(\vec{x},\gamma)\nonumber\\
\hat{B}(\vec{x})=\frac{-i}{2}\varepsilon^{ij}\triangle_{ij}(\vec{x})\nonumber\\
\partial_{i}\hat{C}(\vec{x})=i\triangle_{i}(\vec{x})\nonumber\\
\hat{D}(\vec{x})=T(\vec{x},X)
\end{eqnarray}

and the generator of duality can be written down as
\begin{eqnarray}
\hat{G}=\frac{1}{4\pi}\int d^{2}x\int
d^{2}y\varepsilon^{ij}\triangle_{ij}(\vec{y})\triangle_{k}(\vec{x})\frac{(\vec{x}-\vec{y})}{|\vec{x}-\vec{y}|}-\frac{1}{2\pi}\oint_{\gamma}
d^{2}x^{j}\sum_{a}\varepsilon^{ij}s_{a}\frac{(x_{a}-x)^{i}}{|\vec{x}_{a}-\vec{x}|^{2}}
\end{eqnarray}

As in the previous massless cases, the last term measures a
"linking number", this time between closed paths and points in a
$2$-surface. In other words, it detects wether or not the loop
encloses a given point, taking into account both the sense (clock
our counterclockwise) in which the path surrounds the point, and
the "sign" of the point. This term can be rewritten as the
intersection number between the loop and open paths emanating from
the positive points or ending at the negative ones, in the form

\begin{eqnarray}
\hat{G}=\frac{1}{4\pi}\int d^{2}x\int
d^{2}y\varepsilon^{ij}\triangle(\vec{y})_{ij}\triangle(\vec{x})_{k}\frac{(\vec{x}-\vec{y})^{k}}{|\vec{x}-\vec{y}|}\\
- \int_{\gamma}d x^{i}\oint_{c}d
y^{j}\varepsilon^{ij}\delta^{2}(\vec{x}-\vec{y}).
\end{eqnarray}

\section{CONCLUDING REMARKS}

It is worth noticing that in all the cases we have studied the
generator of duality is a metric independent quantity. In fact, in
the massless cases it is given by the sum of two $BF$ terms in the
$(D-1)$-dimensional space. In the case of self duality (for
instance, in Maxwell theory in $D=4$), instead of $BF$ terms we
should speak of two Chern-Simons terms: one of them is constructed
with the vector potential, the second one with a vector potential
for the electric field. In the massive models, the metric
independence is also clear: the generator comprises terms which
are the integral over the $(D-1)$-dimensional space of
$(D-1)$-forms (see equations (\ref{genmad}) and (\ref{genmd})). This fact explains why
in the geometric representations the generator yields generalized
link invariants, as we found.

The topological character of the duality symmetry has also a
different (but related) manifestation that we discuss briefly. In
reference \cite{algebradual} it was shown that the
�electric-magnetic� duality of Abelian gauge theories induces
a description of their physical phase space in terms of a pair of
non-local observables that are dual in the Kramers-Wannier sense.
The algebra that these observables obey is invariant under spatial
diffeomorphisms. For instance, in the case of Maxwell theory in
four space time dimensions, the dual operators are the Wilson Loop
and the �t Hooft disorder operator [5]. In reference
\cite{piojeanyo} this result was also extended to the case of
massive theories, and to the case of the Maxwell-Chern-Simons
theory and the so called Self-Dual theory, which are dual.

It is interesting to observe that for each model we have studied in this article, there is also a  dual
geometric representation. For instance, in the
Maxwell-Klein-Gordon model in $D=3$, which we realized in the
space of electric loops and points associated to the momentum of
the scalar field, we could equally well choose to work in a space
of magnetic loops and points associated with the scalar field.
Then, to  interpret  the generator of duality we should exchange the
roles of the two terms: the one with generalized loops derivatives
turns into a term that measures linking numbers, and viceversa. However, we do not have yet a simple interpretation of both terms of the generator
in a single geometrical representation. It would also be interesting to study if there is a geometric representation analogous to the ones here considered that allows to realize
the generator of duality for linearized gravity, which was study in reference \cite{henne} as a  generalization of the results of \cite{deser}. These and other related matters are under study.

\end{document}